\documentclass[manuscript]{aastex}

\shorttitle{Lost solar cycle in 18th century}
\shortauthors{Usoskin et al.}

\begin{document}


\title{A solar cycle lost in 1793--1800:\\ Early sunspot observations resolve the old mystery}


\author{Ilya G. Usoskin}
\affil{Sodankyl\"a Geophysical Observatory (Oulu unit), University of Oulu, Finland}
\email{ilya.usoskin@oulu.fi}

\author{Kalevi Mursula}
\affil{Dept. of Physical Sciences, University of Oulu, Finland}

\author{Rainer Arlt}
\affil{Astrophysikalisches Institut Potsdam, An der Sternwarte 16,
 14482 Potsdam, Germany}

\and

\author{Gennady A. Kovaltsov}
\affil{Ioffe Physical-Technical Institute of RAS, 194021 St.Petersburg, Russia}




\begin{abstract}
Because of the lack of reliable sunspot observation, the quality of sunspot number series is poor
 in the late 18th century, leading to the abnormally long solar cycle (1784--1799) before the Dalton minimum.
Using the newly recovered solar drawings by the 18--19th century observers Staudacher and Hamilton,
 we construct the solar butterfly diagram, i.e. the latitudinal
 distribution of sunspots in the 1790's.
The sudden, systematic occurrence of sunspots at high solar latitudes in
 1793--1796 unambiguously shows that a new cycle started in 1793, which was lost
 in traditional Wolf's sunspot series.
This finally confirms the existence of the lost cycle that has been proposed earlier,
 thus resolving an old mystery.
This letter brings the attention of the scientific community to the need of revising the
 sunspot series in the 18th century.
The presence of a new short, asymmetric cycle implies changes and constraints to
 sunspot cycle statistics, solar activity predictions, solar dynamo theories
 as well as for solar-terrestrial relations.
\end{abstract}





\section{Introduction}

Starting from the first telescopic sunspot observations by David and Johannes Fabricius, Galileo Galilei, Thomas Harriot
 and Christoph Scheiner, the 400-year sunspot record is one of the longest directly recorded scientific data series,
 and forms the basis for numerous studies in
 a wide range of research such as, e.g., solar and stellar physics, solar-terrestrial relations, geophysics, and climatology.
During the 400-year interval, sunspots depict a great deal of variability from the extremely quiet period
 of the Maunder minimum \citep{eddy76} to the very active modern time \citep{solanki04}.
The sunspot numbers also form a benchmark data series, upon which virtually all modern models of long-term solar dynamo evolution,
 either theoretical or (semi)empirical, are based.
Accordingly it is important to review the reliability of this series, especially since it contains
 essential uncertainties in the earlier part.

The first sunspot number series was introduced by Rudolf Wolf who observed sunspots from 1848 until 1893,
 and constructed the monthly sunspot numbers since 1749 using archival records and proxy data
 \citep{wolf61}.
Sunspot activity is dominated by the 11-year cyclicity, and the cycles are numbered in
 Wolf's series to start with cycle \#1 in 1755.
When constructing his sunspot series Wolf interpolated over periods of sparse or missing
 sunspot observations using geomagnetic proxy data, thus losing the actual detailed temporal evolution of sunspots
 \citep{hoyt94,hoyt98}.
Sunspot observations were particularly sparse in the 1790's, during solar cycle \#4 which
 became the longest solar cycle in Wolf's reconstruction with an abnormally long declining
 phase (see Fig.~\ref{Fig:wsn}A).
The quality of Wolf's sunspot series during that period has been questioned since long.
Based on independent auroral observations, it was proposed by Elias Loomis already in 1870
 that one small solar cycle may have been completely lost in Wolf's sunspot
 reconstruction in the 1790's \citep{loomis}, being hidden inside the interpolated, exceptionally
 long declining phase of solar cycle \#4.
This extraordinary idea was not accepted at that time.
A century later, possible errors in Wolf's compilation for the late 18th century have been
 emphasized again based on detailed studies of Wolf's sunspot series \citep{gnevyshev48,sonett83JGR}.

Recently, a more extensive and consistent sunspot number series (Fig.~\ref{Fig:wsn}B), the group sunspot numbers (GSN),
 was introduced by \citet{hoyt98}, which increases
 temporal resolution and allows to evaluate the statistical uncertainty of sunspot numbers.
We note that the GSN series is based on a more extensive database than Wolf's series and
 explicitly includes all the data collected by Wolf.
However, it still depicts large data gaps in 1792--1794 (this interval was interpolated in Wolf's series).
Based on a detailed study of the GSN series, \citet{usoskin01,usoskin02} revived Loomis' idea
 by showing that the lost cycle (a new small cycle started in 1793, which was lost
 in the conventional Wolf sunspot series) agrees with both the GSN data (Fig.~\ref{Fig:wsn}B) and
 indirect solar proxies (aurorae) and does not contradict with the cosmogenic isotope data.
The existence of the lost cycle has been disputed by \citet{krivova02} based on data of cosmogenic isotope and sunspot numbers.
However, as argued by \citet{usoskin02,usoskin03}, the lost cycle
 hypothesis does not contradict with sunspots or cosmogenic isotopes
 and is supported by aurorae observations. 
Using time series analysis of sparse sunspot counts or sunspot proxies,
 it is hardly possible to finally verify the existence of the lost solar cycle.
Therefore, the presence of the lost cycle has so far remained as an unresolved issue.

Here we analyze newly restored original solar drawings of the late 18th century to ultimately
 resolve the old mystery and to finally confirm the existence of the lost cycle.

\section{Data and analysis}

\subsection{Positional sunspot data}
\begin{figure}
\begin{center}
\resizebox{8cm}{!}{\includegraphics{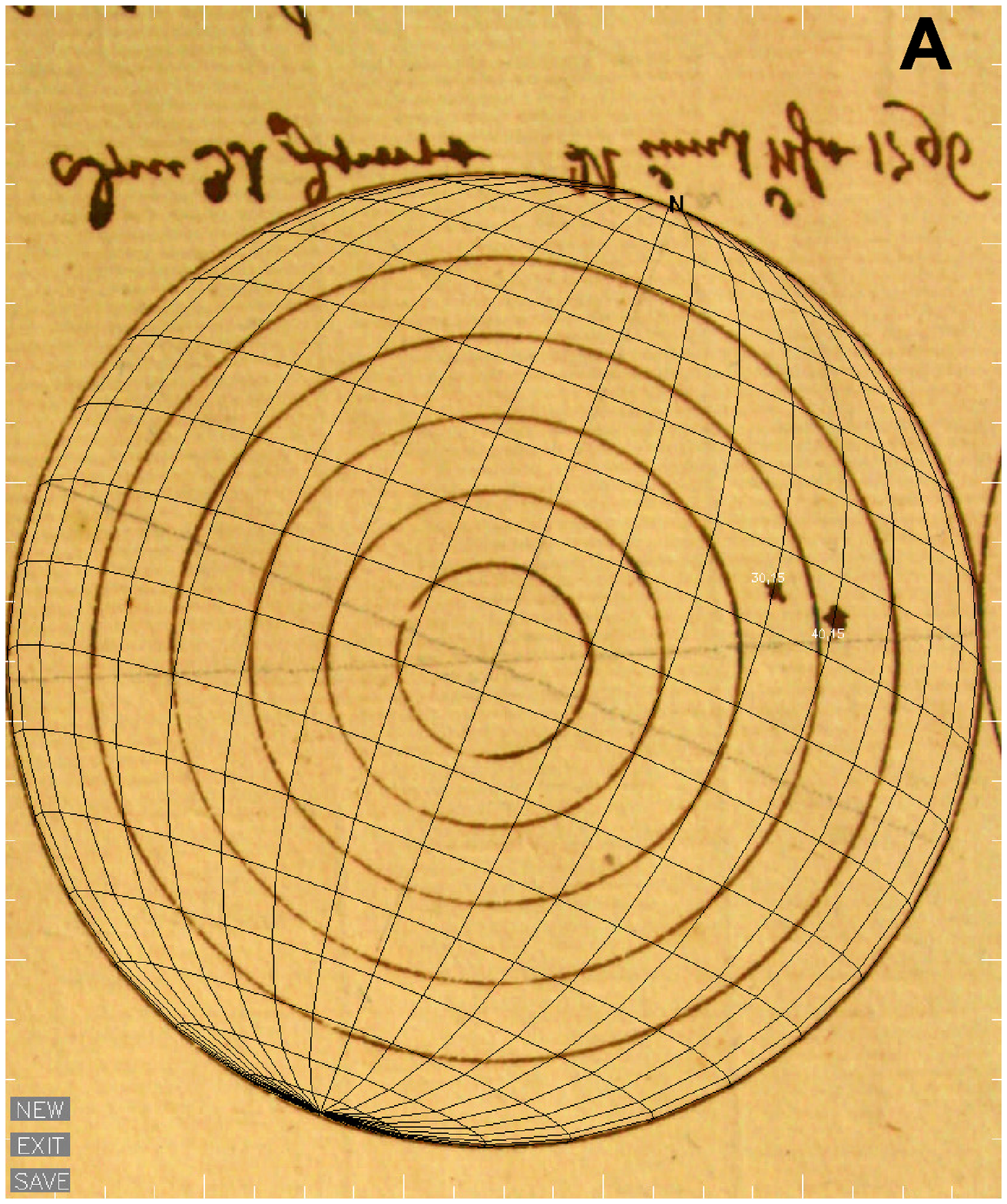}}
\end{center}
\caption{
An example of drawings of sunspots on the solar disc made by Staudacher in N\"urnberg,
 Germany at 1 p.m. local time on January 31, 1796.
The image is mirrored to correspond to the real view to the solar disc.
The heliographical grid has been included during the image processing \citep{arlt09}.
\label{Fig:staud}}
\end{figure}

Most of Wolf's sunspot numbers in 1749-1796 were constructed from observations
 by the German amateur astronomer Johann Staudacher
 who not only counted sunspots but also drew solar images in the second half of the
 18th century (see an example in Fig.~\ref{Fig:staud}).
However, only sunspot counts have so far been used in the sunspot series,
 but the spatial distribution of spots in these drawings
 has not been analyzed earlier.
The first analysis of this data, which covers the lost cycle period in 1790's, has been made only recently \citep{arlt08}
 using Staudacher's original drawings.
Additionally, a few original solar disc drawings made by the
 Irish astronomer James Archibald Hamilton and his assistant since 1795
 have been recently found in the archive of the Armagh Observatory \citep{arlt_AN}.
After the digitization and processing of these two sets of original drawings \citep{arlt08,arlt09,arlt_AN}, the
 location of individual sunspots on the solar disc in the late 18th century has been determined.
This makes it possible to construct the sunspot butterfly diagram for solar cycles \#3 and 4 (Fig.~\ref{Fig:wsn}C), which
 allows us to study the existence of the lost cycle more reliably than based on sunspot counts only.

\subsection{Constructing the butterfly diagram from data with uncertainties}

\begin{figure}
\plotone{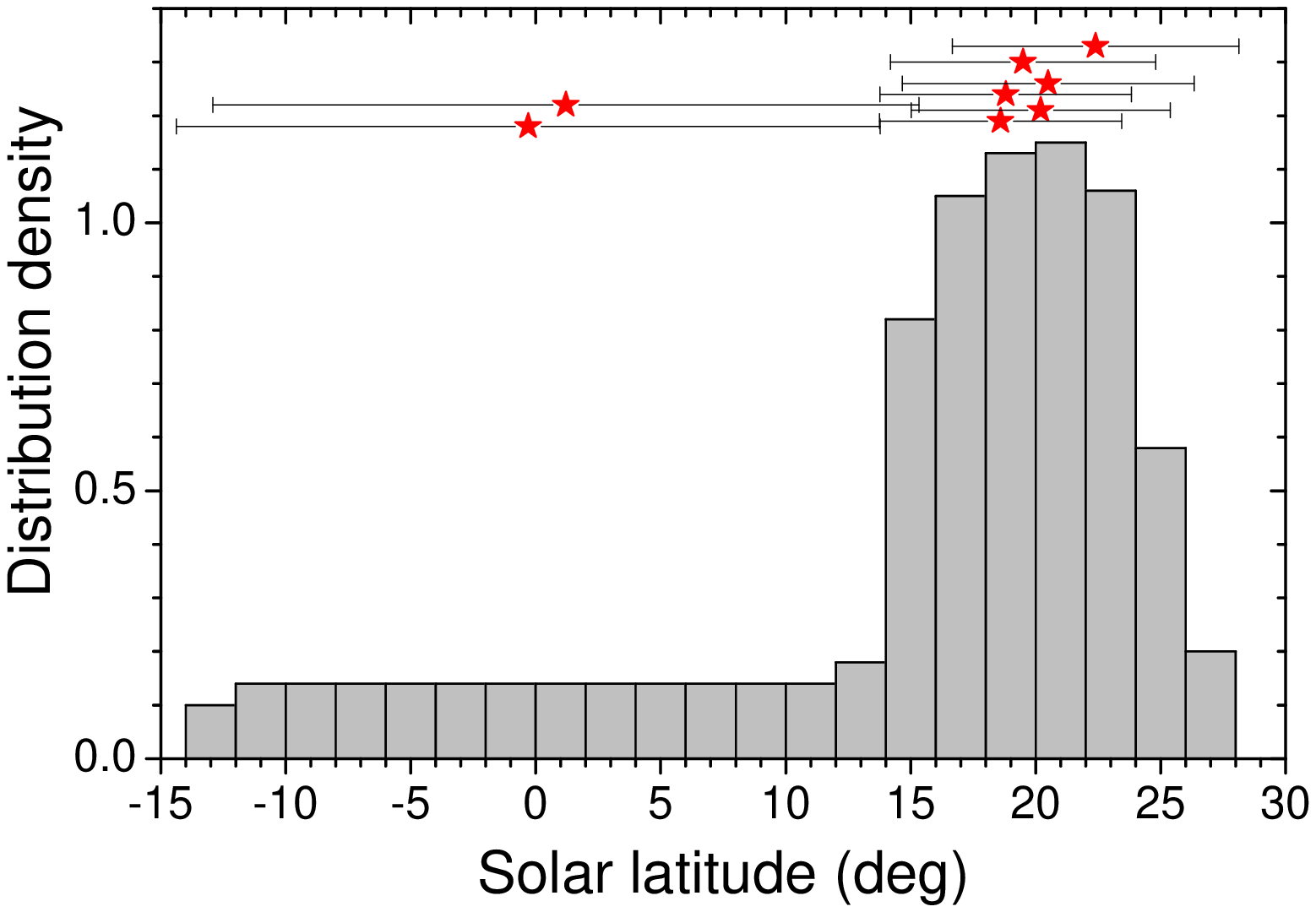}
\caption{An example of the sunspot latitudinal distribution for Jul-Dec 1793,
 with 2$^\circ$ latitudinal bins.
Stars with error bars denote latitudes of individual spots as defined from
 Staudacher's drawings.
The histogram depicts the density of the latitudinal distribution of sunspots
 per 2$^\circ$ bins per half-year computed by including the uncertainties.
\label{Fig:dist}}
\end{figure}

Despite the good quality of original drawings, there is an uncertainty in determining the actual
 latitude for some sunspots (see \cite{arlt09,arlt_AN} for details).
This is related to the limited information on the solar equator in these drawings.
The drawings which are mirrored images of the actual solar disc
 as observed from Earth, cannot be analyzed by an automatic prodecure
 adding the heliographic grid.
Therefore, special efforts have been made to determine the solar equator and
 to place the grid of true solar coordinates for each drawing (see Fig.~\ref{Fig:staud}).
Depending on the information available for each drawing, the uncertainty in defining the solar equator,
 $\Delta Q$, ranges from almost 0$^\circ$ up to a maximum of 15$^\circ$ \citep{arlt09}.
The latitude error of a sunspot, identified to appear at latitude $\lambda$,
 can be defined as
\begin{equation}
\Delta_\lambda = \Delta Q\cdot \sin(\alpha)
\label{eq:delta}
\end{equation}
where $\Delta Q$ is the angular uncertainty of the solar equator in the respective drawing,
 and $\alpha$ is the angular distance between the spot and the solar disc center.
Accordingly, the final uncertainty $\Delta_\lambda$ can range from 0$^\circ$ (precise
 definition of the equator or central location of the spot) up to 15$^\circ$.
We take the uncertain spot location into account
 when constructing the semiannually averaged butterfly diagram as follows.
Let us illustrate the diagram construction for the second half-year (Jul-Dec) of 1793
 (Fig.~\ref{Fig:dist}).
During this period there were only two daily drawings by Staudacher with the total of 8 sunspots:
 two spots on August 6th, which were located close to the limb near the equator,
 and six spots on November 3rd, located near the disc center at higher latitudes.
The uncertainty in definition of the equator was large ($\Delta Q=15^\circ$) for both drawings.
Because of the near-limb location (large $\alpha$) of the first two spots, the error $\Delta_\lambda$
 of latitude definition (Eq.~\ref{eq:delta}) is quite large.
The high-latitude spots of the second drawing are more precisely determined because of the
  central location of the spots.
The latitudinal occurrence of these eight spots and their uncertainties are shown in Fig.~\ref{Fig:dist}
 as stars with error bars.
The true position of a spot is within the latitudinal band $\lambda\pm\Delta_\lambda$,
 where $\Delta_\lambda$ is regarded as an observational error
  and $\lambda$ as the formal center of the latitudinal band.
Accordingly, when constructing the butterfly diagram, we spread the occurrence of each
 spot within this latitudinal band with equal probability (the use of other
 distribution does not affect the result).
Finally, the density of the latitudinal distribution of spots during the analyzed period
 is computed as shown by the histogram in Fig.~\ref{Fig:dist}.
This density is the average number of sunspots occurring per half-year per 2$^\circ$ latitudinal bin.
Each vertical column in the final butterfly diagram shown in Fig.~\ref{Fig:wsn}C is in fact such a histogram
 for the corresponding half-year.

\subsection{Statistical test}

Typically, the sunspots of a new cycle appear at rather high latitudes of about 20--30$^\circ$.
This takes place around the solar cycle minimum.
Later, as the new cycle evolves, the sunspot emergence zone slowly moves towards the solar equator.
This recurrent ``butterfly''-like pattern of sunspot occurrence is known as the {\it Sp\"orer law}
 \citep{maunder04} and is related to the action of the solar dynamo \citep[see, e.g.][]{charbonneau05}.
It is important that the systematic appearance of sunspots at high latitudes unambiguously indicates the beginning
 of a new cycle \citep{waldmeier75} and thus may clearly distinguish between the cycles.

One can see from the reconstructed butterfly diagram (Fig.~\ref{Fig:wsn}C) that the sunspots in 1793--1796
 appeared dominantly at high latitudes, clearly higher than the previous sunspots that belong
 to the late declining phase of the ending solar cycle \#4.
Thus, a new ``butterfly'' wing starts in late 1793, indicating the beginning of the lost cycle.

Since sunspot observations are quite sparse during that period,
 we have performed a thorough statistical test as follows.
The location information of sunspot occurrence on the original drawings during 1793--1796
 (summarized in Table~\ref{Tab1}) allows us to test the existence of the lost cycle.
The observed sunspot latitudes were binned into three categories: low ($<8^\circ$),
 mid- ($8^\circ$--$16^\circ$) and high latitudes ($>16^\circ$), as summarized in
 column 2 of Table~\ref{Tab2}.
We use all available data on latitude distribution of sunspots since 1874 covering solar cycles 12 through 23
 (The combined Royal Greenwich Observatory (1874--1981) and USAF/NOAA (1981--2007)
 Sunspot Data set: http://solarscience.msfc.nasa.gov/greenwch.shtml) as the reference data set.
We tested first if the observed latitude distribution of sunspots (three daily observations
 with low-latitude spots, one with mid-latitude and three with only high-latitude spots, see Table~\ref{Tab2})
 is consistent with a late declining phase (D-scenario, i.e. the period 1793--1796 corresponds
 to the extended declining phase of cycle \#4) or with the early ascending phase (A-scenario, i.e.,
 the period 1793--1796 corresponds to the ascending phase of the lost cycle).
We have selected two subsets from the reference data set: D-subset corresponding to the declining phase which
 covers three last years of solar cycles 12 through 23 and includes in total 11235 days when 33803 sunspot regions
 were observed; and A-subset corresponding to the early ascending phase which covers 3 first years of
 solar cycles 13 through 23 and includes 10433 days when 47096 regions were observed.

First we analyzed the probability to observe sunspot activity of each category on a randomly chosen day.
For example, we found in the D-subset 4290 days when sunspots were observed at low latitudes below 8$^\circ$.
This gives the probability $p=0.38$ (see first line, column 3 in Table~\ref{Tab2}) to observe such a pattern
 on a random day in the late decline phase of a cycle.
Similar probabilities for the other categories in Table~\ref{Tab2} have been computed in the same way.
Next we tested whether the observed low-latitude spot occurrence (three out of seven
 daily observations) corresponds to declining/ascending phase scenario.
The corresponding probability to observe $n$ events (low-latitude spots) during
 $m$ trials (observational days) is given as
\begin{equation}
P_m^n = p^n\cdot (1-p)^{m-n}\cdot C_m^n,
\label{eq:P}
\end{equation}
where $p$ is the probability to observe the event at a single trial, and
 $C_m^n$ is the number of possible combinations.
We assume here that the results of individual trials are independent on each other,
 which is justified by the long separation between observational days.
Thus, the probability to observed three low-latitude spots during seven random days
 is $P_7^3=0.27$ and 0.07 for D- and A-hypotheses, respectively.
The corresponding probabilities are given in the first row, columns 5--6 of Table~\ref{Tab2}.
The occurrence of three days with low latitude activity is quite probable for both declining
 and ascending phases.
Thus, this criterion cannot distinguish between the two cases.
The observed mid-latitude spot occurrence (one out of seven daily observations) is also
 consistent with both D- and A-scenarios.
The corresponding confidence levels (0.06 and 0.22, respectively, see the second row, columns 5--6 of Table~\ref{Tab2})
 do not allow to select between the two hypotheses.
Next we tested the observed high-latitude spot occurrence (three out of seven
 observations) in the D/A-scenarios (the corresponding probabilities are given in the third row of Table~\ref{Tab2}).
The occurrence of three days with high-latitude activity is highly improbable
 during a late declining phase (D-scenario).
Thus, the hypothesis of the extended cycle \#4
  is rejected at the level of $5\cdot 10^{-4}$.
The A-scenario is well consistent (confidence 0.26) with the data.
Thus, the observed high-latitude sunspot occurrence clearly confirms the existence of the
 lost cycle.

We also noticed that sunspots tend to appear in Northern hemisphere (13 out of 16 observed sunspots
 appeared in the Northern hemisphere).
Despite the rather small number of observations, the statistical significance of asymmetry is
 quite good (confidence level 99\%), i.e. it can be obtained by chance with the probability of
 only 0.01, in a purely symmetric distribution.
Nevertheless, more data are needed to clearly evaluate the asymmetry.

Thus, a statistical test of the sunspot occurrence during 1793--1796 confirms that:
\begin{itemize}
\item
The sunspot occurrence in 1793--1796 contradicts with a typical latitudinal
 pattern in the late declining phase of a normal solar cycle (at the significance level of $5\cdot 10^{-4}$).

\item
The sunspot occurrence in 1793--1796 is consistent with a typical
 ascending phase of the solar cycle, confirming the start of the lost solar cycle.
We note that it has been shown earlier \citep{usoskin03}, using the group sunspot number,
 that the sunspot number distribution during 1792--1793 was statistically
 similar to that in the minimum years of a normal solar cycle,
 but significantly different from that in the declining phase.

\item
The observed asymmetric occurrence of sunspots during the lost cycle
 is statistically significant (at the significance level of 0.01).

\end{itemize}
Therefore, the sunspot butterfly diagram (Fig.~\ref{Fig:wsn}C) unambiguously proves the existence
 of the lost cycle in the late 18th century,
 verifying the earlier evidence based on sunspot numbers \citep{usoskin01,usoskin03}
 and aurorae borealis \citep{loomis,usoskin02}.

\section{Discussion and Conclusions}

\begin{figure}
\begin{center}
\epsscale{.80}
\plotone{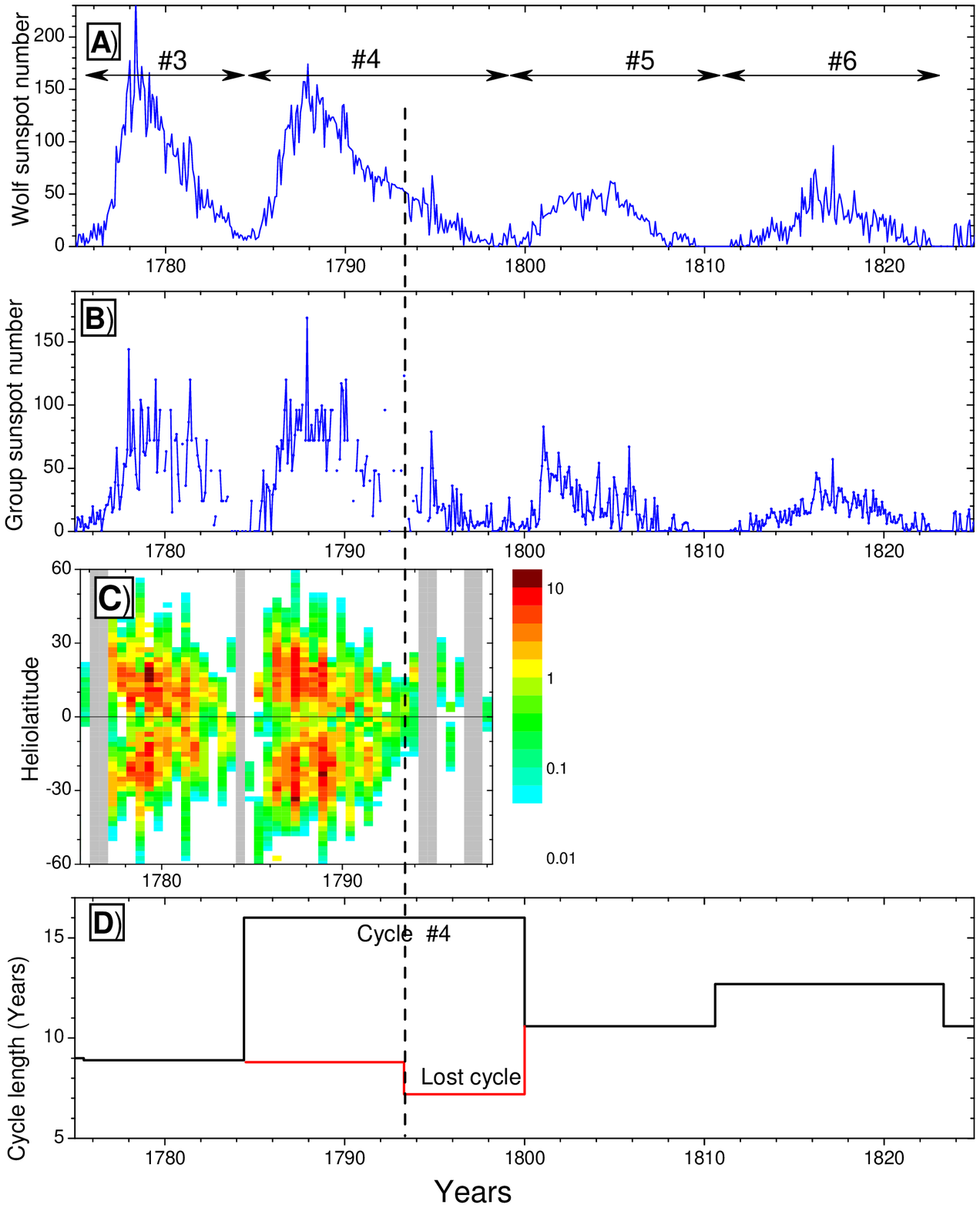}
\end{center}
\caption{
Sunspot activity in late 18th and early 19th century.
The start of the lost cycle (late 1793) is denoted by a vertical dashed line.
{\bf A:} Monthly Wolf sunspot numbers with conventional solar cycle numbers shown on the top.
{\bf B}: Monthly group sunspot numbers.
{\bf C}: The newly reconstructed sunspot butterfly diagram, which takes into account the uncertainties
 of the estimate sunspot latitudes.
The color scale on the right gives the density (in year$^{-1}$deg$^{-1}$) of sunspots
 in latitude-time bins (one bin covers 2$^\circ$ in latitude and 6 months in time).
Grey bars indicate that no latitudinal information is available for the corresponding half-year.
{\bf D}: Lengths of solar cycles.
The conventional lengths using the group sunspot numbers is shown by the black line, while
 the red line depicts the cycle lengths after including the lost cycle.
\label{Fig:wsn}}
\end{figure}

An additional cycle in the 1790's changes cycle numbering before
 the Dalton minimum, thus verifying the validity of the {\it Gnevyshev-Ohl} rule of sunspot cycle pairing \citep{gnevyshev48}
  and the related 22-year periodicity \citep{mursula01} in sunspot activity throughout the whole 400-year interval.
Another important consequence of the lost cycle is that, instead of one abnormally long cycle \#4 (min-to-min length
 $\approx$15.5 years according to GSN \citep{usoskin02}) there are two shorter cycles of about 9 and 7 years (see Fig.~\ref{Fig:wsn}D).
Note also that some physical dynamo models even predict the existence of cycles of small amplitude
and short duration near a grand minimum \citep{kuker99}.
The cycle \#4 (1784--1799 in GSN) with its abnormally long duration dominates empirical
 studies of relations, e.g., between cycle length and amplitude.
Replacing an abnormally long cycle \#4 by one fairly typical and one small short cycle
 changes empirical relations based on cycle length statistics.
This will affect, e.g., predictions of future solar activity by statistical
 or dynamo-based models \citep{dicke78,dikpati06,braisa09}, and some important
 solar-terrestrial relations \citep{friis91,kelly92}.

The lost cycle starting in 1793 depicts notable hemispheric asymmetry with most sunspots of the new cycle occurring
 in the northern solar hemisphere (Fig.~\ref{Fig:wsn}C).
This asymmetry is statistically significant at the confidence level of 99\%.
A similar, highly asymmetric sunspot distribution existed
 during the Maunder minimum of sunspot activity in the second half of the 17th century \citep{soon03,nesme}.
However, the sunspots during the Maunder minimum occurred preferably in the southern solar hemisphere \citep{sokoloff94},
 i.e., opposite to the asymmetry of the lost cycle.
This shows that the asymmetry is not constant, contrary to some earlier models
 involving the fossil solar magnetic field \citep{bravo95,boruta96,usoskin00}.
Interestingly, this change in hemispheric asymmetry between the Maunder and Dalton minimum
 is in agreement with an earlier, independent observation, based on long-term geomagnetic activity, that the
 north-south asymmetry oscillates at the period of about 200-250 years \citep{mursula_z01}.

Concluding, the newly recovered spatial distribution of sunspots of the late 18th
 century conclusively confirms the existence of a new solar cycle in 1793--1800,
 which has been lost under the preceding, abnormally long cycle compiled by Rudolf Wolf
 when interpolating over the sparse sunspot observations of the late 1790's.
This letter brings the attention of the scientific community to the need of revising the
 sunspot series in the 18th century and the solar cycle statistics.
This emphasizes the need to search for new, yet unrecovered, solar data to restore
 details of solar activity evolution in the past \citep[e.g.,][]{vaquero07}.
The new cycle revises the long-held sunspot number series, restoring
 its cyclic evolution in the 18th century and modifying the statistics of all solar cycle
 related parameters.
The northern dominance of sunspot activity during the lost cycle suggests that hemispheric
 asymmetry is typical during grand minima of solar activity, and gives independent support for a systematic,
 century-scale oscillating pattern of solar hemispheric asymmetry.
These results have immediate practical and theoretical consequences, e.g., to predicting
 future solar activity and understanding the action of the solar dynamo.



\acknowledgments

We are grateful to Dr. John Butler from Armagh Observatory for his
 help with finding the old notes of Hamilton's data.
Support from the Academy of Finland and
 Finnish Academy of Sciences and Letters (V\"ais\"al\"a Foundation) are acknowledged.

\clearpage



%
\begin{table}
\caption{
Sunspot occurrence during the lost cycle with dates, number ($N$) and latitude range
 ($\lambda$) of the observed spots on each day.
\label{Tab1}}
\begin{tabular}{l|c|c}
\hline
\hline
Date and Observer & $N$ & $\lambda$ \\
\hline
1793.08.06 Staudacher& 2 & $<3^\circ$ \\
1793.11.03 Staudacher& 6 & $>18^\circ$ N \\
1795.02.19 Staudacher& 3 & $>20^\circ$ N \\
1795.10.08--15 Hamilton & 2 & 15$^\circ$ S and 6$^\circ$ N \\
1795.11.02--03 Hamilton & 1 & $5.5^\circ$ S\\
1796.01.31\tablenotemark{a} Staudacher & 2 & $>16^\circ$ N \\
\hline
\end{tabular}
\tablenotetext{a}{Shown in Fig.~\ref{Fig:staud}.}
\end{table}
\begin{table}
\caption{
Number of days ($n$) with observed sunspots in 1793--1796, sunspot latitude ranges,
 probability ($p$) of sunspots to be found on a randomly selected day, and the cumulative probability
 ($P$), calculated for D- and A-scenaria using all data since 1874.
 \label{Tab2}}
\begin{tabular}{c|c|cc|cc}
\hline
\hline
$n$ & Latitude range & \multicolumn{2}{c|}{probability $p$} & \multicolumn{2}{c}{cumulative probability $P$}\\
 & & D-scenario & A-scenario & D-scenario & A-scenario\\
\hline
3 & Low latitude ($<8^\circ$) & 0.38 & 0.1 & 0.27 & 0.07\\
1 & Mid-latitude (8--16$^\circ$) & 0.48 & 0.32 & 0.06 & 0.22\\
3 & High latitude ($>16^\circ$) & 0.026 & 0.52 & 0.0005 & 0.26\\
\hline
\end{tabular}
\end{table}

\clearpage

\end{document}